\documentclass[notitlepage,superscriptaddress,showpacs,a4paper,nobalancelastpage,twocolumn,aps,prx]{revtex4-1}
\RequirePackage[T1]{fontenc}
\RequirePackage{times} % favourite for note.

\usepackage{amsfonts}
\usepackage{subfigure}
\usepackage{amsmath}
\usepackage{amssymb}
\usepackage{graphicx,epstopdf}
\usepackage{appendix}
\usepackage{array}
\usepackage{color}
\usepackage{my_symbols}
\usepackage{CJK}
{\par}

\begin{document}
\begin{CJK*}{UTF8}{gbsn} % Use default fonts from CJK (see below)
\title{A spin wave diode}
%\author{Jin Lan}
\author{Jin Lan (兰金)}
\thanks{These authors contributed equally.}
\affiliation{Department of Physics and State Key Laboratory of Surface Physics, Fudan University, Shanghai 200433, China}
%\author{Weichao Yu}
\author{Weichao Yu (余伟超)}
\thanks{These authors contributed equally.}
\affiliation{Department of Physics and State Key Laboratory of Surface Physics, Fudan University, Shanghai 200433, China}
\author{Ruqian Wu}
\affiliation{Department of Physics and State Key Laboratory of Surface Physics, Fudan University, Shanghai 200433, China}
\affiliation{Department of Physics and Astronomy, University of California, Irvine, California 92697-4575, USA}
%\author{Jiang Xiao}
\author{Jiang Xiao (萧江)}
\email[Corresponding author:~]{xiaojiang@fudan.edu.cn}
\affiliation{Department of Physics and State Key Laboratory of Surface Physics, Fudan University, Shanghai 200433, China}
\affiliation{Collaborative Innovation Center of Advanced Microstructures, Fudan University, Shanghai 200433, China}

\begin{abstract}
A diode, a device allowing unidirectional signal transmission, is a fundamental element of logic structures and lies in the heart of modern information systems. Spin wave or magnon, representing a collective quasi-particle excitation of the magnetic order in magnetic materials, is a promising candidate of information carrier for the next generation energy-saving technologies. Here we propose a scalable and reprogrammable pure spin wave logic hardware architecture using domain walls and surface anisotropy stripes as waveguides on a single magnetic wafer. We demonstrate theoretically the design principle of the simplest logic component, a spin wave diode, utilizing the chiral bound states in a magnetic domain wall with Dzyaloshiskii-Moriya interaction, and confirm its performance through micromagnetic simulations. Our findings open a new vista for realizing different types of pure spin wave logic components and finally achieving an energy-efficient and hardware-reprogrammable spin wave computer. 

%The counter propagating bound states along magnetic domain walls with Dzyaloshiskii-Moriya interaction are found to be chiral and spatially separated. Using the domain wall with these chiral bound states as a waveguide, a design of spin wave diode is proposed and confirmed by micromagnetic simulations. The forward direction of the diode is reconfigurable by shifting the domain wall position via current or magnon-induced spin-transfer torque. A reprogrammable spin wave logic hardware architecture is proposed based on the domain wall waveguide and the surface anisotropy waveguide on a magnetic wafer.
\end{abstract}

\maketitle
\end{CJK*}

%{\bf A diode, a device allowing unidirectional signal transmission, is a fundamental element of logic structures and lies in the heart of modern information systems. \cite{tocci_thinfilm_1995,li_thermal_2004,chang_solid-state_2006,liang_acoustic_2009,liang_acoustic_2010,borlenghi_designing_2014} Spin wave or magnon, \cite{ashcroft_solid_1985,gurevich_magnetization_1996, stancil_spin_2009} representing a collective quasi-particle excitation of the magnetic order in magnetic materials, is a promising candidate of information carrier for the next generation energy-saving technologies. \cite{serga_yig_2010,kruglyak_magnonics_2010,khitun_magnonic_2010,lenk_building_2011} Here we propose a scalable and reprogrammable pure spin wave logic hardware architecture using domain walls and surface anisotropy stripes as waveguides on a single magnetic wafer. We demonstrate theoretically the design principle of the simplest logic component, a spin wave diode, utilizing the chiral bound states in a magnetic domain wall with Dzyaloshiskii-Moriya interaction, \cite{dzyaloshinsky_thermodynamic_1958, moriya_anisotropic_1960} and confirm its performance through micromagnetic simulations. Our findings open a new vista for realizing different types of pure spin wave logic components and finally achieving an energy-efficient and hardware-reprogrammable spin wave computer. } 
%
%\bigskip

\section{Introduction}

In the post silicon era, Moore's law is not sustainable, partly due to the power consumption caused by the Joule heating from electric current. To avoid the unmanageable power dissipation, people have been trying to use various (quasi-) particles other than electrons as information carrier, such as photon in photonics, \cite{obrien_photonic_2009} electron spin in spintronics, \cite{pulizzi_spintronics_2012} phonon in phononics, \cite{li_phononics:_2012,maldovan_sound_2013} and spin wave in magnonics. \cite{serga_yig_2010,kruglyak_magnonics_2010,khitun_magnonic_2010,lenk_building_2011} Among these efforts, the magnonics, which can be realized in insulators, is particularly interesting mainly due to its energy-saving benefit because spin waves produce no Joule heating. Since both spin waves and magnetic memory are associated with the re-ordering of magnetic moments, it is possible and natural to integrate both logic and storage operations through pure spin wave information processing without the need of other information architectures. New magnonics hardware architecture design as we propose below allow magnonics to be realized on a single magnetic thin film --- a magnetic wafer using its ``soft'' magnetic structures. Such an integrated spin wave circuit is reprogrammable by re-patterning the magnetic texture. This is in contrast to most present day electronic technologies that use the ``hard'' physical structures consisting of several materials and layers. 

Controlling the information transmission direction is a basic feature of all information processing systems, as implemented in various diode structures. In addition to the classical electric diode using p-n junction, there are optical diode, \cite{tocci_thinfilm_1995} heat diode, \cite{li_thermal_2004,chang_solid-state_2006} acoustic wave (phonon) diode, \cite{liang_acoustic_2009,liang_acoustic_2010}and spin-Seebeck diode etc. \cite{borlenghi_designing_2014} In this paper, we propose a design of the spin wave diode utilizing the spatial separation of the spin wave bound states caused by the Dzyaloshinskii-Moriya interaction (DMI) \cite{dzyaloshinsky_thermodynamic_1958, moriya_anisotropic_1960} within a magnetic domain wall. The DMI is an antisymmetric exchange coupling induced by spin-orbit interaction in magnetic materials with broken inversion symmetry, either in bulk lattice or at the interface. The functionality of this reprogrammable spin wave diode is confirmed by micromagnetic simulations. 

\section{Magnetic wafer based spin wave architecture}

%=============================
\begin{figure*}[t]
\centering
\subfigure[an EASA wire and its crossing with a domain wall wire]{\includegraphics[height=4.5cm]{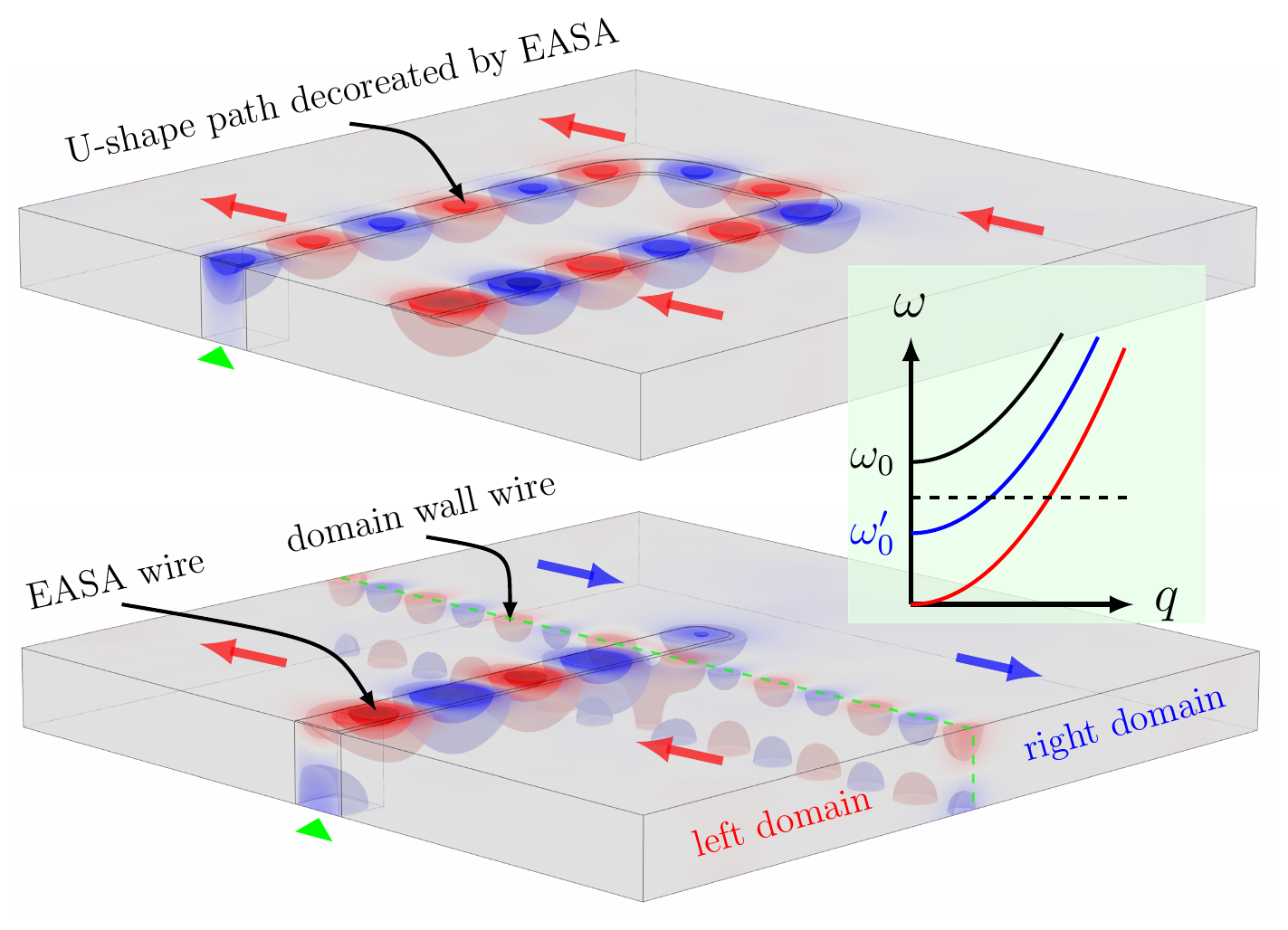}} \hspace{0.1cm}
\subfigure[Domain wall circuit]{ \includegraphics[height=4.5cm]{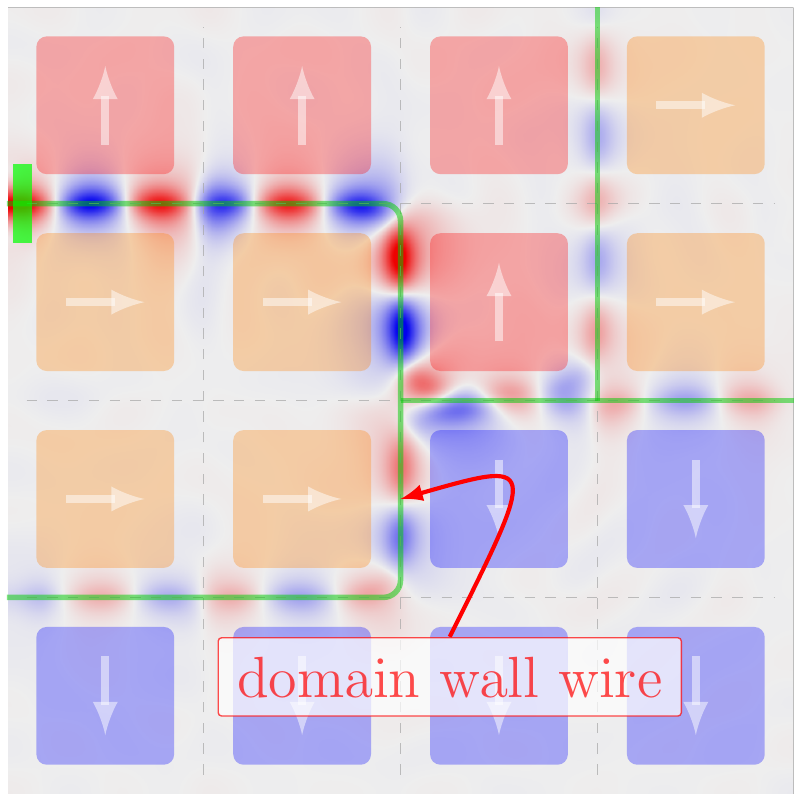}}\hspace{0.1cm}
\subfigure[EASA circuit]{ \includegraphics[height=4.5cm]{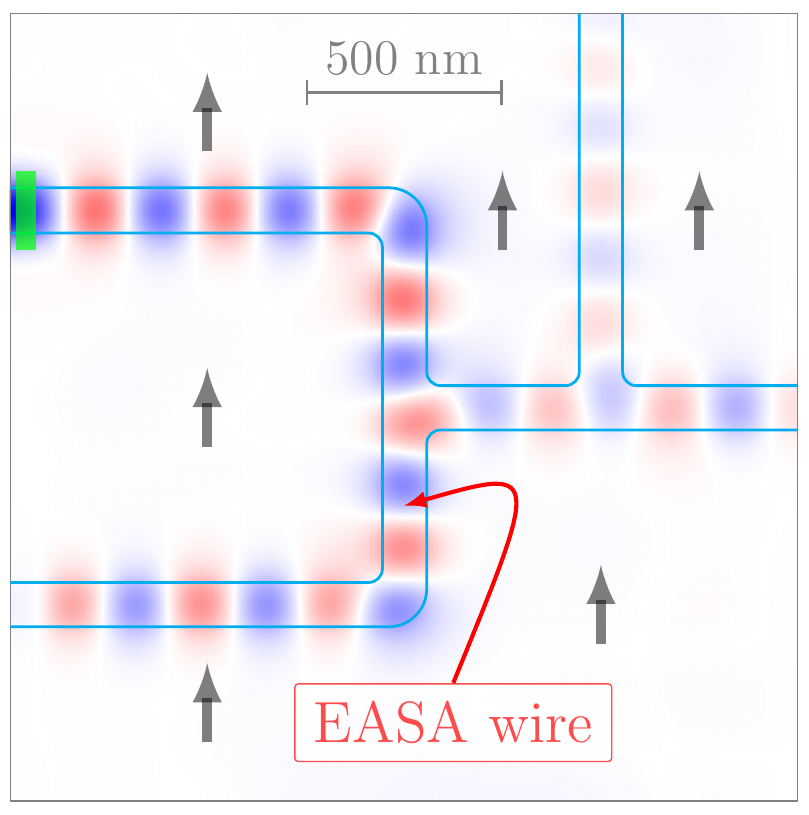}}
\caption{The domain wall circuit and EASA circuit. (a), upper: simulated EASA induced surface spin wave mode propagates along a U-shape path decorated by EASA; lower: simulated spin wave interconnection from an EASA wire to a domain wall wire; both images are simulated on an homogeneous magnetic wafer of size 1600nm$\times$1600nm$\times$150nm with spin wave excitation at the position indicated by the triangle; inset: the dispersion relation for the bulk spin wave (black), EASA surface mode (blue), and domain wall bound state (red); the excitation frequency is indicated by the dashed line lying below the bulk gap and above the surface wave gap. (b), simulated spin wave propagation in a domain wall circuit made of the domain wall wires printed on a magnetic wafer with a biaxial anisotropy and four possible domain orientations, and wafer size 2000nm$\times$2000nm$\times$10nm. (c), simulated EASA induced surface spin wave transporting through an EASA circuit in a homogeneous magnetic wafer of size 2000nm$\times$2000nm$\times$30nm. }
\label{fig:swcircuits}
\end{figure*}
%=============================

%\noindent {\bf Magnetic wafer based spin wave architecture.} 
To construct spin wave logic components on a 2-dimensional magnetic wafer, we first need waveguides or wires that can transport spin waves. A magnetic domain wall can be a natural waveguide using the domain wall bound state as carrier as demonstrated numerically by Garcia-Sanchez \etal \cite{garcia-sanchez_narrow_2015}. Another type of waveguide is to utilize the surface spin wave mode induced by the easy-axis surface anisotropy (EASA), \cite{xiao_spin-wave_2012,zhou_current-induced_2013} with which the surface spins tend to point in the surface normal direction. \cite{gurevich_magnetization_1996} To construct an EASA wire, the surface of the magnetic wafer is decorated by EASA along the wire path for the surface mode to propagate. Such decoration can be either a capping layer of other materials or simply a process that modifies the original surface structure. The penetration depth of this EASA surface mode is inversely proportional to the strength of EASA, \ie the stronger the EASA, the shallower the penetration. \cite{xiao_spin-wave_2012,zhou_current-induced_2013} Both types of spin wave waveguides do not have hard structure on the magnetic wafer, especially the domain wall wire, which can be even moved to another position. The domain wall wire and EASA wire can work simply because the dispersions for the domain wall bound state and EASA surface state has either no gap or smaller gap than the bulk spin waves. Within domains, ignoring the dipolar coupling, the bulk spin wave dispersion is: $\omega_{\rm Bulk} = \omega_0 + A_{\rm ex}q^2$ (the top black curve in \Figure{fig:swcircuits}(a) inset), where $\omega_0$ is the spin wave gap in the bulk, $A_{\rm ex}$ is the exchange coupling constant, and $q$ is the in-plane wave vector. The domain wall bound state is gapless and its dispersion is $\omega_{\rm BS} = A_{\rm ex} q^2$ (the bottom red curve in \Figure{fig:swcircuits}(a) inset) \cite{dodd_solitons_1982, yan_all-magnonic_2011}. The dispersion of the EASA surface spin wave is approximately $\omega_{\rm EASA} = \omega'_0 + A_{\rm ex}q^2$ with $\omega'_0 < \omega_0$ (middle blue curve in \Figure{fig:swcircuits}(a) inset). \cite{xiao_spin-wave_2012} The spin wave modes whose frequency $\omega$ lies within the range $\omega_0' < \omega < \omega_0$ can only propagate in the domain wall wire and the EASA wire, but not in the bulk. Through micromagnetic simulations, we demonstrate in \Figure{fig:swcircuits}(a) the transport of spin waves in a U-shape EASA wire (upper) and across an interconnection (lower) between an EASA wire and a domain wall wire. 

Making use of the spin wave wires, an integrated spin wave circuit can be imprinted onto a magnetic wafer. As an example, for a magnetic wafer with biaxial anisotropy along two perpendicular axes, we may create an artificial chessboard-like pattern such that in each square the magnetic order can point in any of the four possible orientations as shown in \Figure{fig:swcircuits}(b). The domain walls are pinned along the grid lines, which can be carved notches on the wafer surface. Such chessboard-like magnetic structure can be used as 2-dimensional memory similar to the 1-dimensional racetrack memory. \cite{parkin_magnetic_2008} However, instead of using magnetic domains to store information, we propose to use domain walls as spin wave wires, which can be interconnected to form a domain wall circuit. Not only can the domain wall circuit be imprinted in any pattern, but also be rewritten by re-orienting the magnetization direction of each square. Therefore, this type of chessboard domain wall circuit is extremely flexible to construct a large number of different spin wave circuits. Simpler circuits can be realized in a wafer with uniaxial anisotropy. It is also possible to use EASA wires to build spin wave circuit: upon a magnetic wafer, EASA wires can be imprinted into any pattern of circuits by simply decorating the surface of the circuits by EASA as shown in \Figure{fig:swcircuits}(c). A more general design may use a hybrid circuit that contains both domain wall and EASA wires. 

%\bigskip

\section{Spin wave diode}

\begin{figure*}[t]
\centering
\subfigure[Domain wall without DMI]{\includegraphics[width=0.22\textwidth]{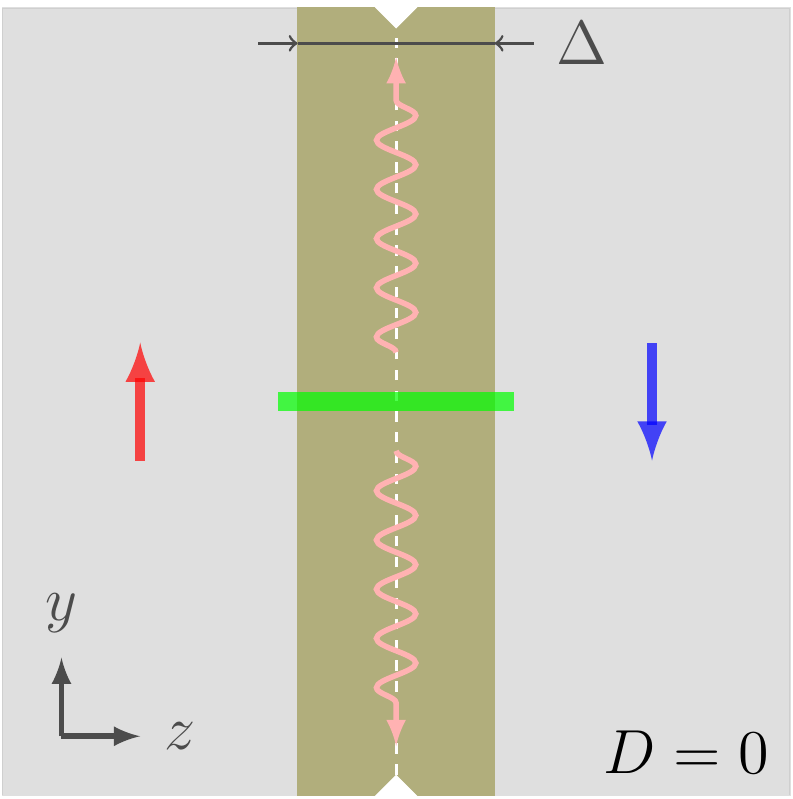}} \hspace{0.1cm}
\subfigure[Domain wall with DMI]{\includegraphics[width=0.22\textwidth]{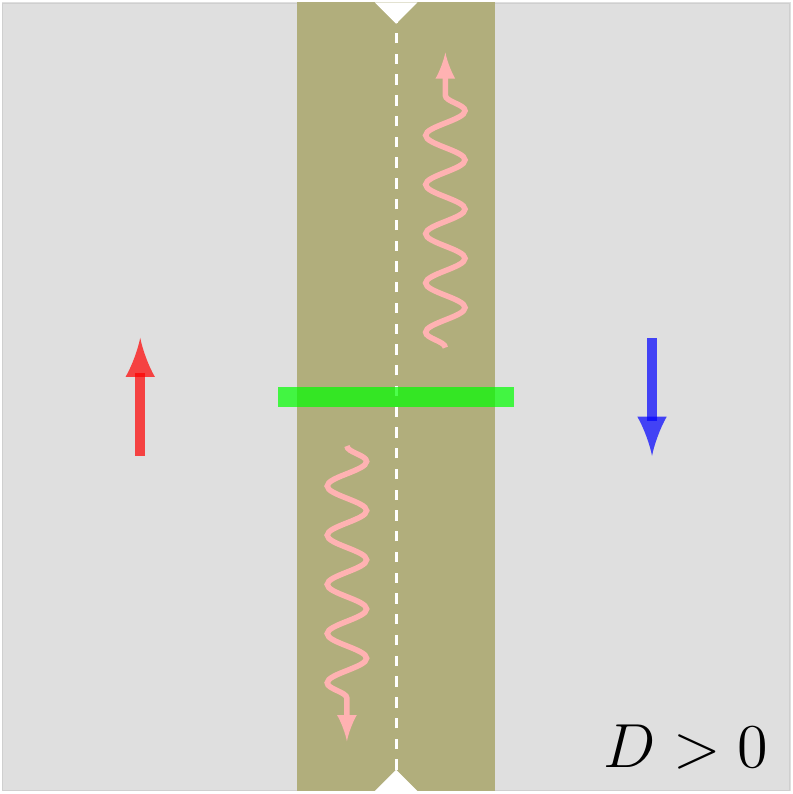}} \hspace{0.1cm}
\subfigure[Diode - forward direction]{\includegraphics[width=0.22\textwidth]{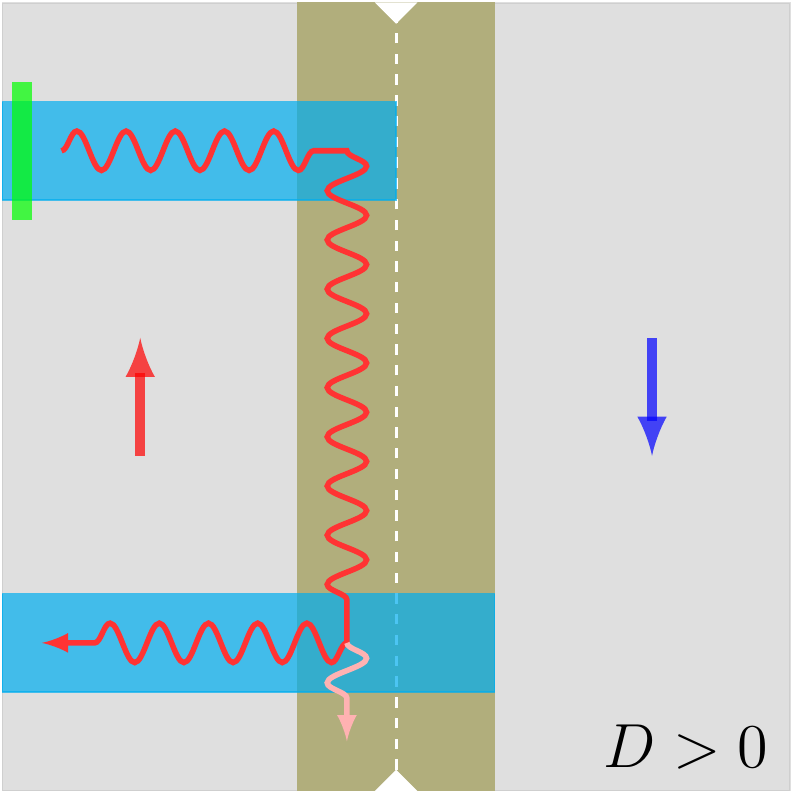}}\hspace{0.1cm}
\subfigure[Diode - reverse direction]{\includegraphics[width=0.22\textwidth]{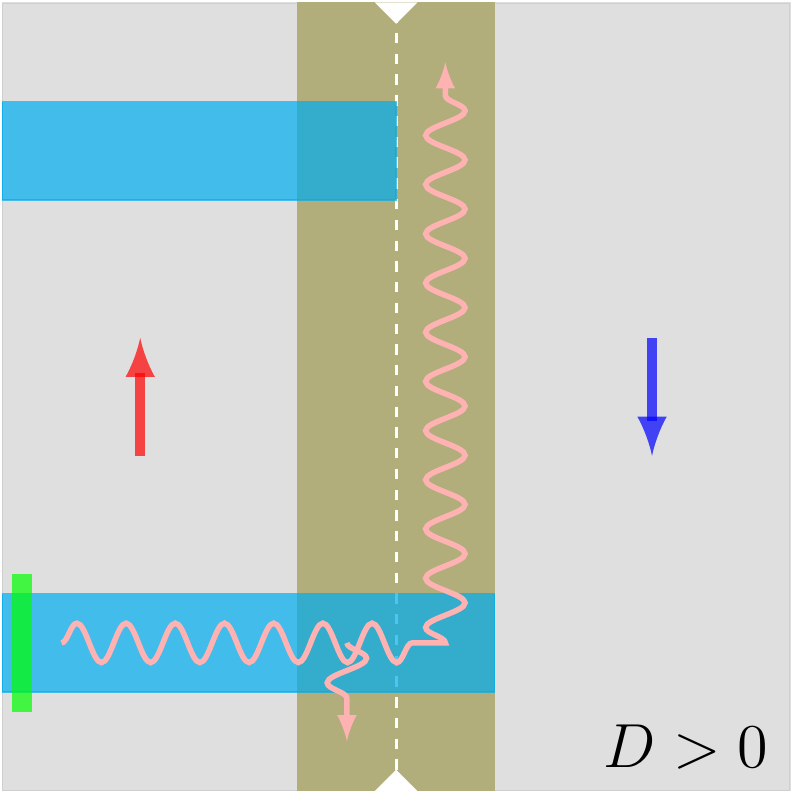}}
\subfigure[Domain wall without DMI]{\includegraphics[width=0.22\textwidth]{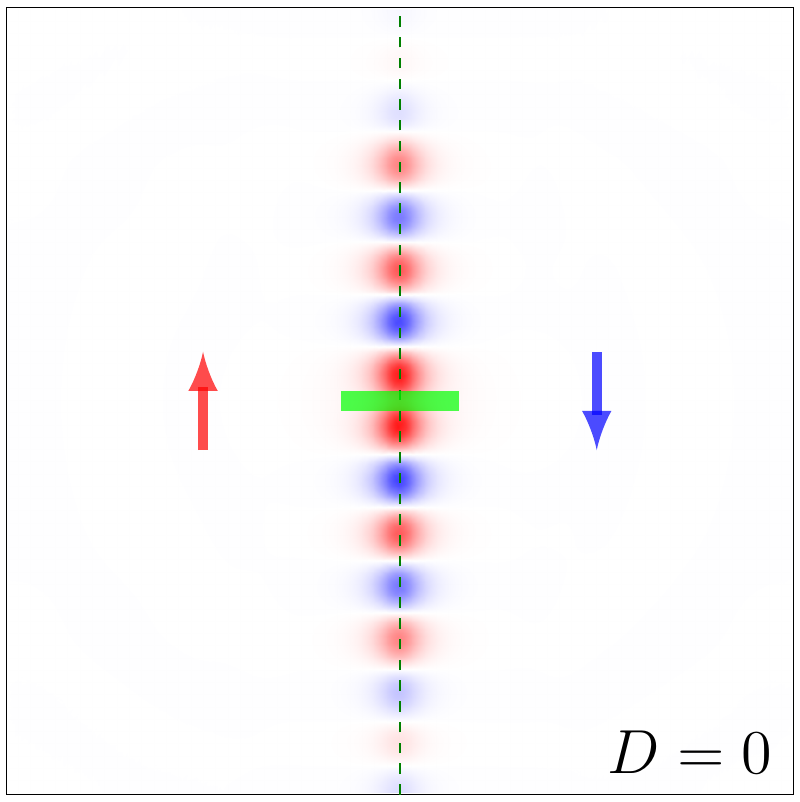}} \hspace{0.1cm}
\subfigure[Domain wall with DMI]{\includegraphics[width=0.22\textwidth]{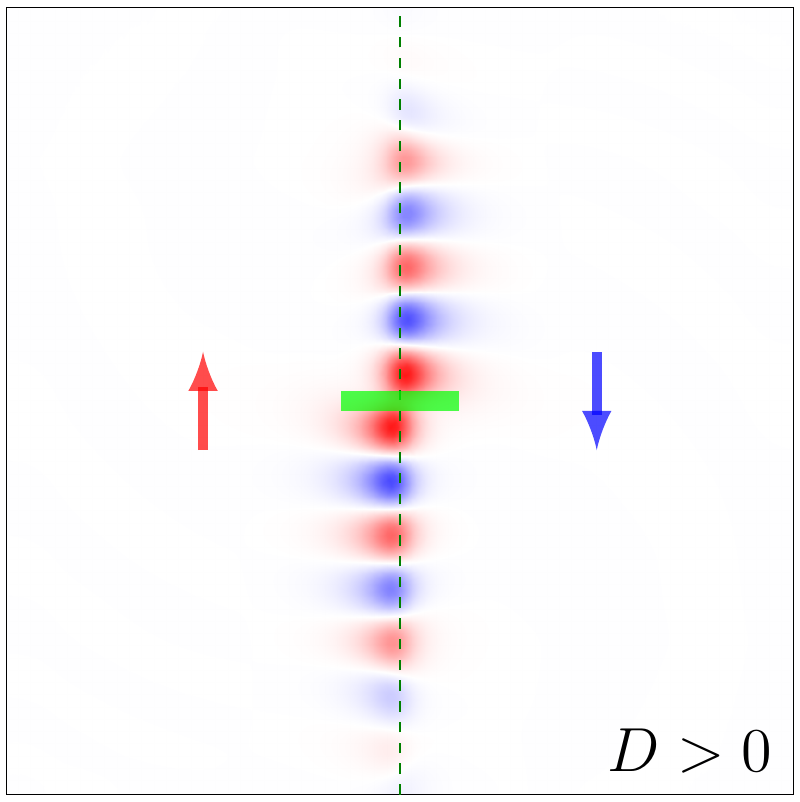}} \hspace{0.1cm}
\subfigure[Diode - forward direction]{\includegraphics[width=0.22\textwidth]{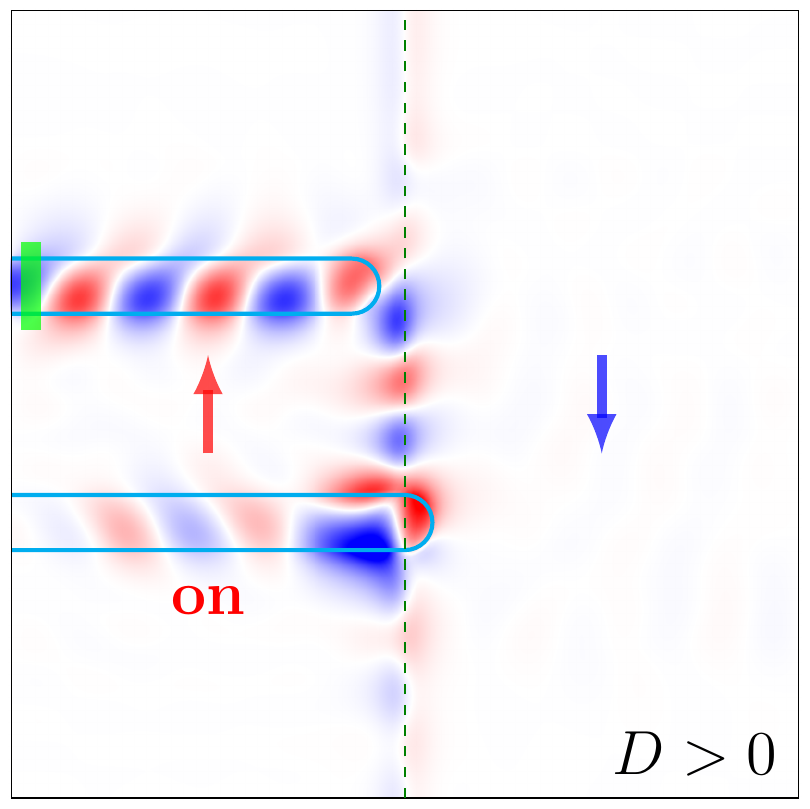}}\hspace{0.1cm}
\subfigure[Diode - reverse direction]{\includegraphics[width=0.22\textwidth]{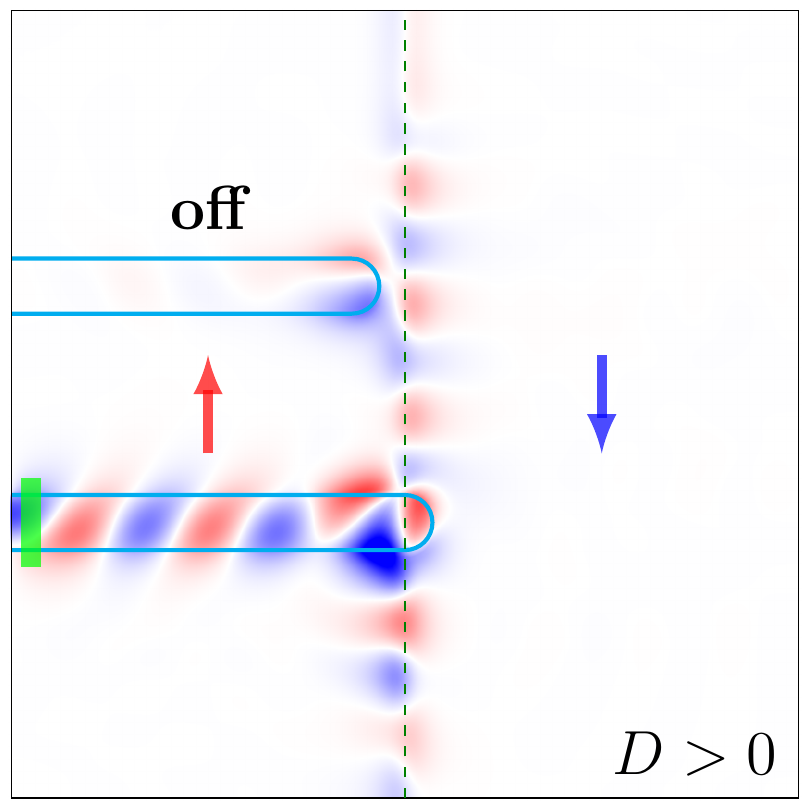}} 
\caption{The design of the diode and the simulations. Top row: (a), Domain wall wire without DMI: the bound spin wave state propagates in both directions identically, where the domain wall region is shaded with darker brown color and is pined at the kink position; (b), Domain wall wire with DMI: the bound states propagating to upward/downward are spatially shifted to the right/left side of the wall; (c), Spin wave diode - forward direction: spin wave transmits from the upper to the lower terminal. (d), Spin wave diode - reverse direction: the spin wave is blocked from lower to upper terminal. The wavy lines denote the route of spin wave propagation. Bottom row: Numerical micromagnetic simulations of the domain wall wire (color map of $m_z$): (e), without DMI, the bound spin wave state travels symmetrically, (f), with DMI, the bound spin wave states become chiral; and spin wave diode: (g), the forward direction, (h), the reverse direction. In all panels, the little green bar indicates the spin wave injection location where an oscillating magnetic field is applied.}
\label{fig:swd}
\end{figure*}

%\noindent {\bf Spin wave diode.} 
To demonstrate the concept of our design of a spin wave diode, we use the magnetic wafer that has uniaxial anisotropy along $y$ direction and contains two magnetic domains with a Bloch domain wall in between as shown in \Figure{fig:swd}(a), where the magnetization in the left/right domain points in $\pm y$ direction, respectively. The magnetization within the wall evolves from $y$ to $-y$ by rotates out of plane in the $x$ direction.   When there is no DMI, the bound spin wave state propagates identically to both $\pm y$ directions along the domain wall wire (see \Figure{fig:swd}(a)). However, the presence of DMI effectively applies a magnetic field in the domain wall region as shown below, and the bound spin wave states that propagate in opposite directions are spatially separated towards two edges of the wire as illustrated in \Figure{fig:swd}(b).

%=============================
%=============================

\subsection{Design principle} 

This spatial separation feature allows us to design the spin wave diode as shown in \Figure{fig:swd}(c, d). The two terminals patterned on the left side of the domain wall (indicated by light blue rectangles) are made of EASA wires. The upper terminal only overlaps with the left half of domain wall, while lower terminal covers the whole domain wall width. The diode works as the following: In the forward direction, the spin wave is injected from the upper terminal as shown in \Figure{fig:swd}(c). The two terminals are connected via the left half of the domain wall wire because the down-going spin wave is localized on the left half of the wall. However, in the reverse direction when the spin wave is injected from the lower terminal, the up-going spin wave is localized on the right half of the domain wall and hence cannot reach the upper terminal as shown in \Figure{fig:swd}(d). This unidirectional transport of spin waves is clearly the diode effect. Note that the spin wave diode is a pure magnetic structure with no physical structure.

The operation of the spin wave diode is confirmed by numerical micromagnetic simulations using a yttrium iron garnet (YIG) thin film of thickness 30nm, with parameters given in the Methods section. \Figure{fig:swd}(e/f) shows the spatial shift of the up-going spin wave bound state in the domain wall wire under the influence of DMI. \Figure{fig:swd}(g/h) shows the diode effect for the forward/reverse spin wave propagations, a behavior that confirms our analyses above. The power loss in the forward direction is about 11.6 dB (power decreased by 14 times), while the power loss in the reverse direction is about 24 dB (power decreased by 254 times), therefore the power loss in the reverse direction is much larger than the forward direction. In these simulations, we artificially set high damping above and below the diode terminals to eliminate the boundary effects. 

%\bigskip

\subsection{Theory} 

%\noindent {\bf Theory. } 
Obviously, the crucial ingredient for realizing the spin wave diode effect is the spatial separation of the spin wave bound states caused by the DMI in the domain wall. To prove this, we adopt the bulk form of DM energy $E_{\rm DMI} = D\mb\cdot(\nabla\times\mb)$, for which Bloch domain wall is favored against N\'{e}el wall. We can understand the chiral feature of the bound states in a semiclassical way by transforming the equation of motion for spin wave dynamics into an effective Schr\"{o}dinger equation. \cite{yan_all-magnonic_2011} The spin wave dynamics is governed by the Landau-Lifshitz-Gilbert (LLG) equation:
%----------------------
\begin{equation}
\dbm = - \gamma\mb\times\bH_{\rm eff} + \alpha \mb\times\dbm,
\label{eqn:LLG}
\end{equation}
%----------------------
where $\mb(\br,t)$ is the unit vector in the direction of the magnetization, $\alpha$ is the Gilbert damping parameter, and $\gamma\bH_{\rm eff} = Km_y\hyy+A_{\rm ex}\nabla^2\mb+D\nabla\times\mb$ is the effective magnetic field acting on $\mb$ due to contributions from anisotropy (along $\hby$), exchange, and DMI. In the absence of DMI, the domain wall width is determined as $\Delta = \sqrt{A_{\rm ex}/K}$, the same for the Bloch and N\'{e}el domain wall. When DMI is included, the domain wall twists slightly for the N\'{e}el wall \cite{wang_magnon-driven_2015}, but remains unchanged for the Bloch wall. Let $\hbm_0 = (\sin\theta_0\cos\phi_0,\sin\theta_0\sin\phi_0,\cos\theta_0)$ be the static magnetic texture of the domain wall along $\hbz$, where $\theta_0(z)$ and $\phi_0(z)$ are the polar and azimuthal angle of $\hbm_0(z)$ with respect to $\hzz$ axis. Let $\delta\mb=m_\theta\hbe_{\theta}+m_\phi\hbe_{\phi}$ be the spin wave excitation on top of the static $\hbm_0$, where $\hbe_{\theta}, \hbe_{\phi}\perp\hbm_0$ are the two transverse directions to $\hbm_0$. As for the dynamics of $\delta\mb$, previous studies show that the effect of the inhomogeneous magnetization texture on $\delta\mb$ can be represented by a scalar potential \cite{yan_all-magnonic_2011}, while the effect of DMI can be represented by a vector potential. \cite{onose_observation_2010}

%=============================
\begin{figure}[t]
\includegraphics[width=0.45\textwidth]{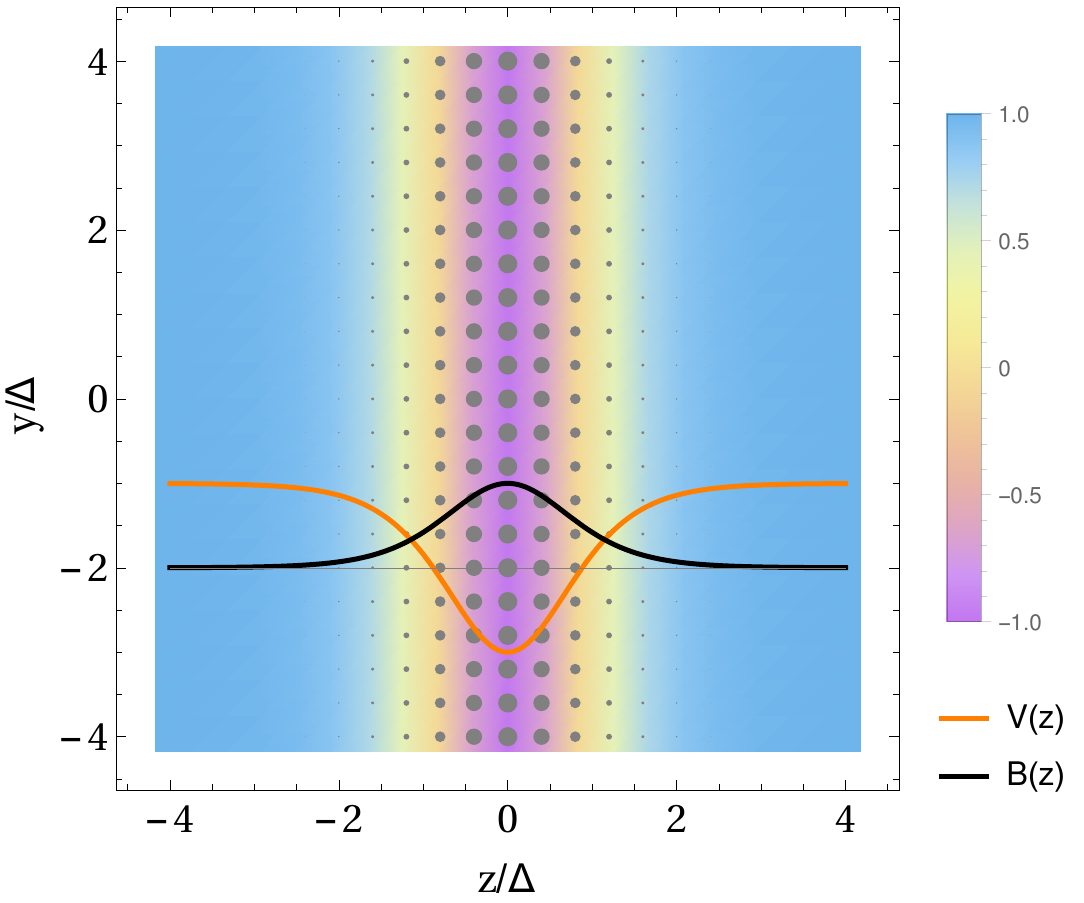}
\caption{The spatial profile for the effective scalar potential and magnetic field. The spatial distribution of the scalar potential $V(z)$ and the magnetic field $B(z)$ in $y$-$z$ plane. The density map is the scalar potential and the disk size represents the magnitude of the magnetic field pointing in $x$ direction. }
\label{fig:VB}
\end{figure}
%=============================

The static structure of a Bloch domain wall is unmodified by the DMI and still takes the Walker profile: $\theta_0(z) = \pi/2, \phi_0(z) = \pi/2+ 2\arctan[\exp(z/\Delta)]$, where the magnetization within the wall rotating out of $y$-$z$ plane in certain direction depending on the sign of $D$. \cite{chen_novel_2013} By redefining $\psi = m_\theta-im_\phi$, the LLG equation (\ref{eqn:LLG}) governing the dynamics of $\delta\mb$ can be recasted into an effective Schr\"{o}dinger equation for $\psi$:
%----------------------
\begin{equation}
\label{eqn:SE}
i\hbar{\partial\ov\partial t}\psi
= \midb{{1\ov 2m^*}\smlb{\hbp - {q\ov c}\bA}^2 + V} \psi,
\end{equation}
%----------------------
where the effective mass $m^* = \hbar/2A_{\rm ex}$, the momentum operator $\hbp = -i\hbar\nabla$, the scalar potential $V = -\hbar K \cos2\phi_0$, 
%$ = \hbar K[1-2\mbox{sech}^2(z/\Delta)]$
and the vector potential $\bA = (Dm^*c/q)\sin\phi_0 \hzz$, which corresponds to an effective magnetic field perpendicular to the $y$-$z$ plane: $\bB = \nabla\times\bA = -(Dm^*c/q) \phi'_0\cos\phi_0 \hxx$.  
%$ = -(Dm^*c/q\Delta) \cos^2\phi_0 \hxx$. 
Therefore, the behavior of the spin wave in a Bloch domain wall structure is equivalent to the motion of a charged particle with mass $m^*$ and charge $q$ in a potential well $V(z)$ and magnetic field $\bB(z)\|\hxx$. The spatial profiles for the potential $V$ and  magnetic field $\bB$ are plotted in \Figure{fig:VB}. The potential well $V(z)$ is a special one not only due to its reflectionlessness, but also for the existence of a bound state at the bottom of the potential well with zero energy in $z$-direction.  

Based on the effective Schr\"{o}dinger equation (\ref{eqn:SE}), we may understand the transport behavior of the bound states within domain walls semiclassically: a) for $D = 0$, the bound state only feels the potential well $V(z)$ with a vanishing magnetic field $\bB = 0$, so it is confined in $z$ direction within the well and travels along the domain wall in $\pm y$ directions symmetrically as shown in \Figure{fig:swd}(a); b) however, for $D > 0$, the effective magnetic field $\bB$ pointing in the $\hxx$ direction (perpendicular to the thin film) is non-zero and maximizes at the domain wall center (see \Figure{fig:VB}). Consequently the spin wave moving upwards to $+y$ (downwards to $-y$) bends to the right (left) due to the effective Lorentz force (see \Figure{fig:swd}(b)). A head-to-head N\'{e}el domain wall can also be stabilized by the DMI, but it does not have the spatial separation behavior because the corresponding vector potential $\bA$ is proportional to $D^2$, much weaker than a Bloch wall.  

%\bigskip

\subsection{Reprogrammability} 

%=============================
\begin{figure}[t]
\centering
\subfigure[Reverse direction]{\includegraphics[width=0.23\textwidth]{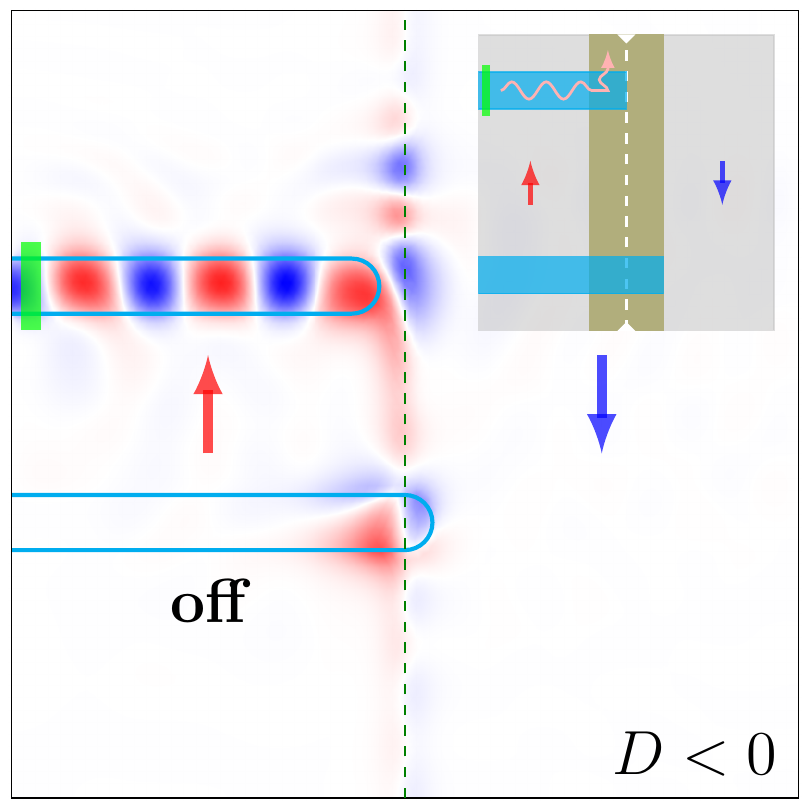}}\hspace{0.3cm}
\subfigure[Forward direction]{\includegraphics[width=0.23\textwidth]{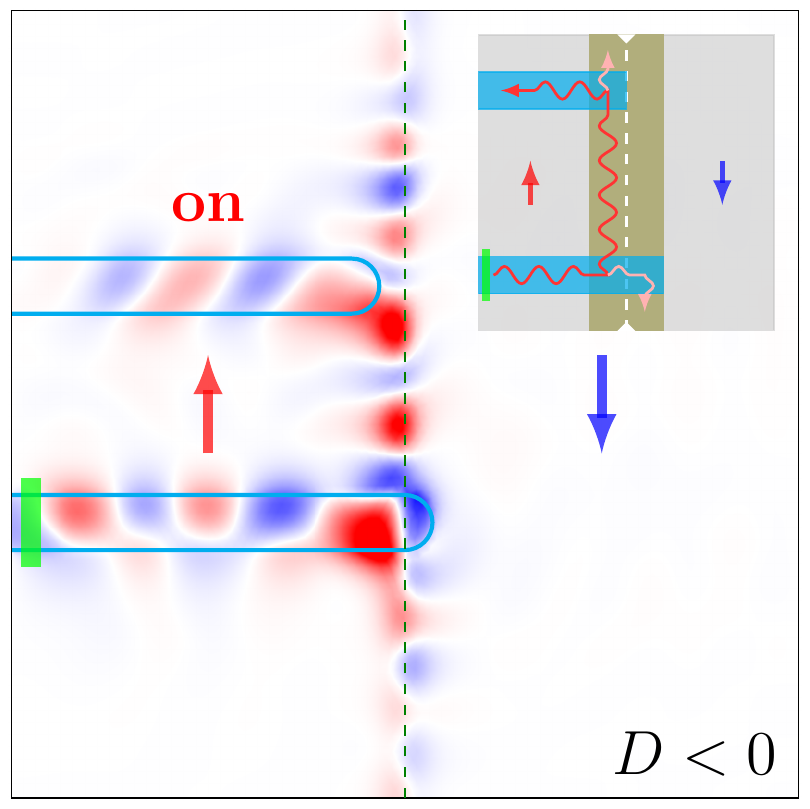}} 
\caption{Reprogramming the diode by changing the sign of the DMI parameter. The forward direction of the spin wave diode for $D < 0$ is opposite to that for $D > 0$ in \Figure{fig:swd}(g, h). The insets on the top right are the spin wave propagation diagrams. }
%\subfigure[$D>0$]{\includegraphics[width=0.15\textwidth]{Dp}} \hspace{0.5cm}
%\subfigure[$D<0$]{\includegraphics[width=0.15\textwidth]{Dn}} 
%\caption{\capf{\textbf{Reprogramming the diode by changing the sign of the DMI parameter.} \textbf{a}, $D > 0$, the forward direction is from left to right. \textbf{b}, $D < 0$, the forward direction is from right to left. }}
\label{fig:swdD}
\end{figure}
%=============================

%\noindent {\bf Reprogrammability.} 
%{\it Changing the sign of $D$ by electric field and thus changing the polarity of the diode}. 
If DMI originates from the inversion symmetry breaking at the surface in magnetic thin films, the spin wave diode works just as well, but for a head-to-head N\'{e}el domain wall instead. The advantage of interfacial DMI is its tunability by external electric field, which can tune the magnitude and even the sign of the DMI parameter $D$.  \cite{heron_deterministic_2014,nawaoka_voltage_2015} In our system, the $D$ can be tuned by applying a gating voltage throughout the whole thin film. When $D$ changes sign ($D < 0$), the direction of the effective magnetic field $\bB$ also changes from $\hxx$ to $-\hxx$, so the spin wave moving to $+y$ ($-y$) now bends to the left (right), opposite to that in \Figure{fig:swd}(b,f). Consequently, the forward direction of the spin wave diode also changes as shown in \Figure{fig:swdD}. In fact, by continuously tuning the value and sign of $D$ using a perpendicular electric field, the spin wave diode can have i) L$\ra$R transmission only ($D > 0$), ii) R$\ra$L transmission only ($D < 0$), iii) L$\leftrightarrow$R transmission ($D \simeq 0$). We demonstrate the real time switching of the three stages of transmission by continuously tuning parameter $D$ in a movie in supplementary materials.   

%=============================
\begin{figure*}[t]
\centering
\subfigure[]{\includegraphics[width=0.19\textwidth]{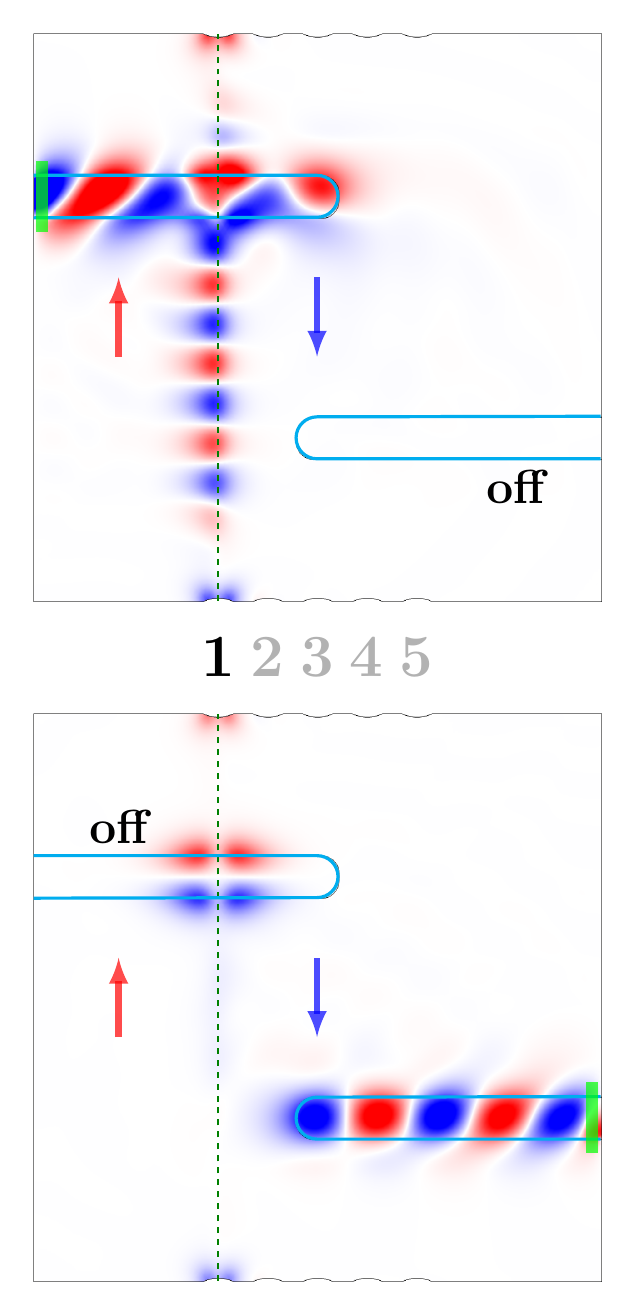}}
\subfigure[]{\includegraphics[width=0.19\textwidth]{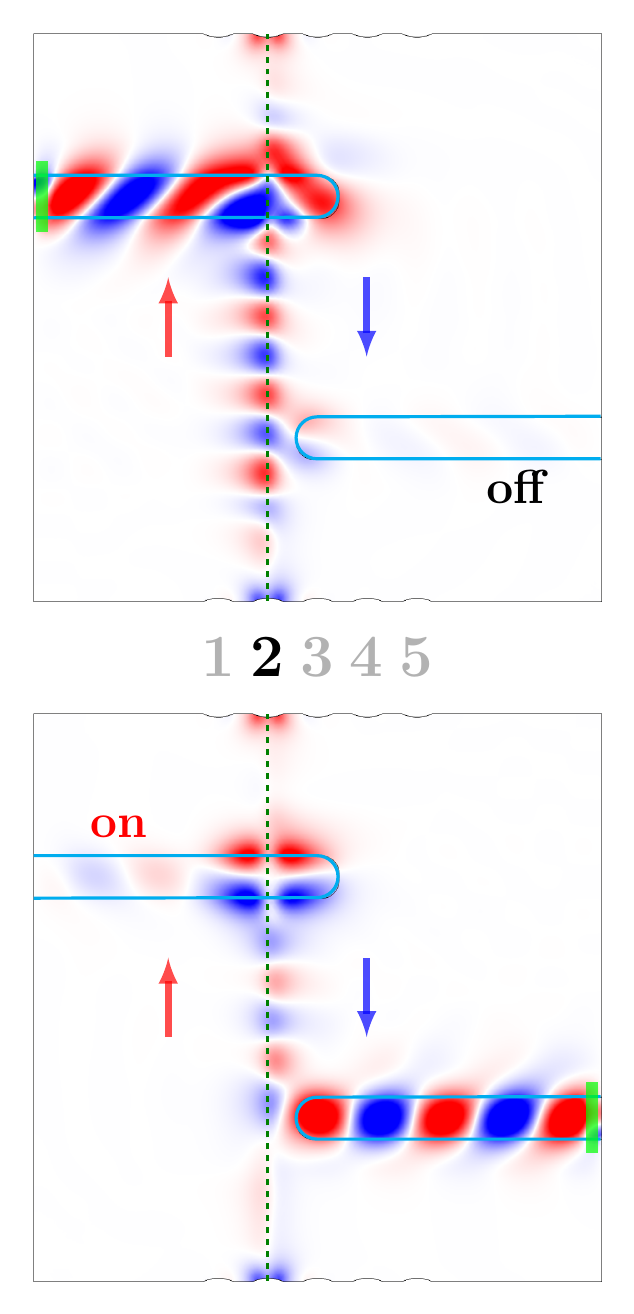}}
\subfigure[]{\includegraphics[width=0.19\textwidth]{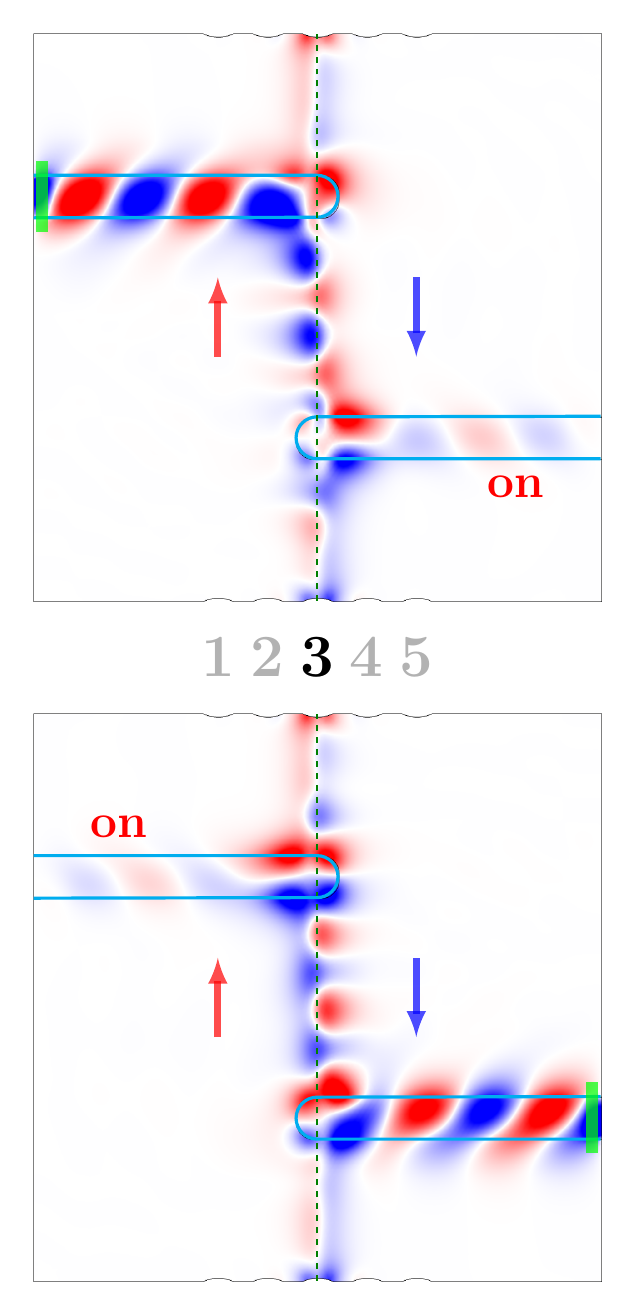}}
\subfigure[]{\includegraphics[width=0.19\textwidth]{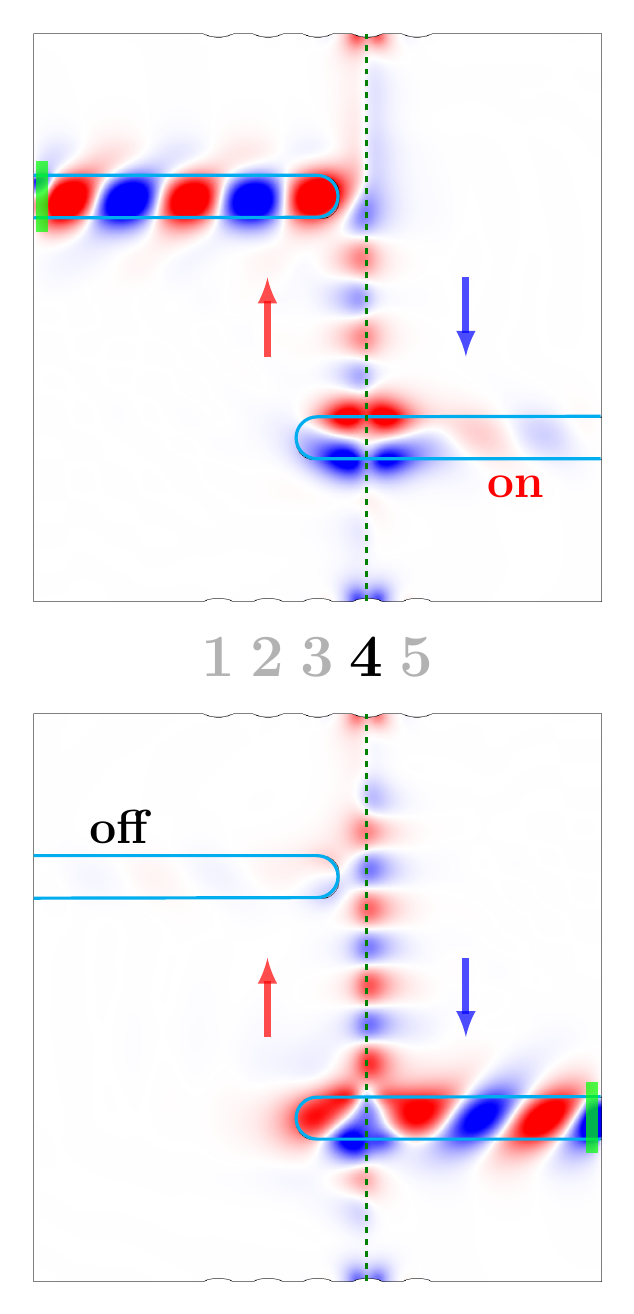}}
\subfigure[]{\includegraphics[width=0.19\textwidth]{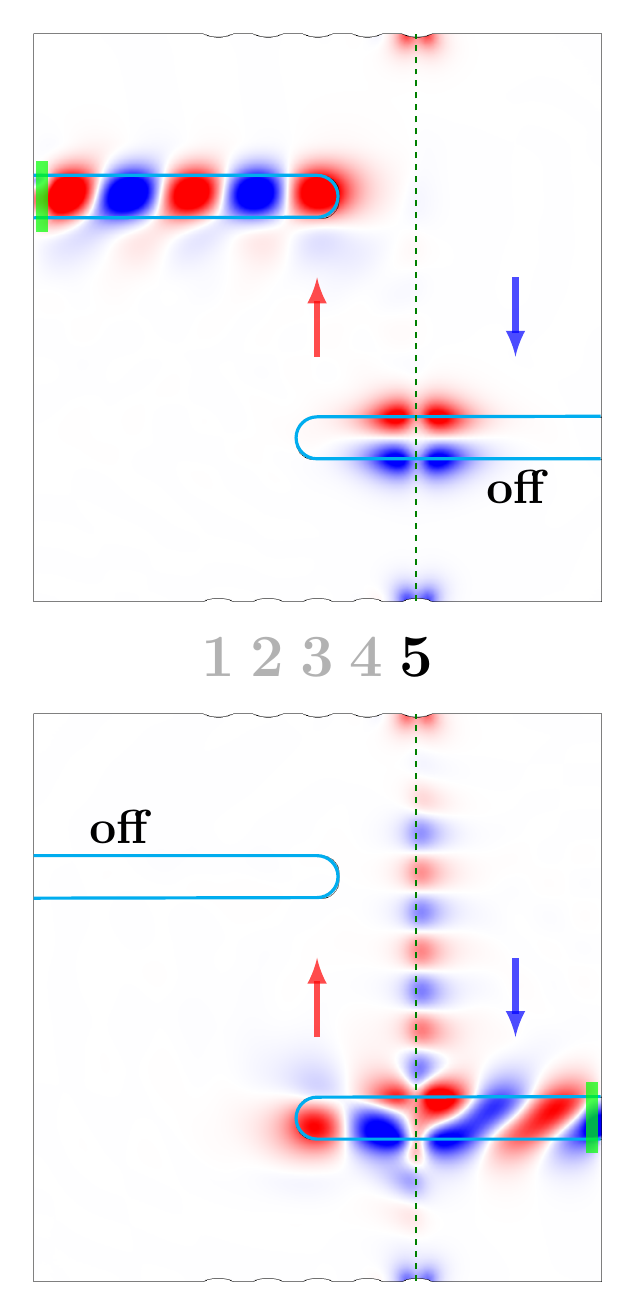}} 
\caption{Reprogramming the diode by moving domain wall position. The functionality is tuned by the domain wall position (from position 1 to 5, domain wall center is indicated by the dashed line). Spin wave is injected from left (right) terminal for the upper (lower) row. (a), position 1, both way off. (b), position 2, diode with forward direction from right to left. (c), position 3, two way on. (d), position 4, diode with forward direction from left to right, opposite to position 2. (e), position 5, both way off, similar to position 1. The kinks are to pin the domain walls. Simulated on a 2-dimensional film of size 1600nm $\times$ 1600nm with an additional perpendicular anisotropy in the terminal area to mimic the EASA in 3-dimensional samples. (See the movie in the supplementary materials for simulated real time switching among these functions by current-induced domain wall motion.) }
\label{fig:swdp}
\end{figure*}
%=============================

%{\it Moving the domain wall to change the polarity of the diode. } 
Another way of tunning the functionality of the spin wave diode is to shift the domain wall position, which can be realized in various ways, for instance, by applying magnetic field, or via current-induced spin-transfer torque, or purely by spin wave with magnonic spin-transfer torque. For this purpose, we design a two-side spin wave diode as shown in \Figure{fig:swdp}, where the terminals are connected to the domain wall from the opposite sides instead of the same side as in \Figure{fig:swd}. \Figure{fig:swdp} shows the functionality of the device for five different domain wall positions pinned by the five kinks. Depending on the domain wall position, the two terminals can either be completely disconnected, manifest the diode effect, or be connected in both ways. Therefore, by repositioning the domain wall, the device function can be easily modified. In the supplementary materials, we demonstrate the real time change of the spin wave transmission properties due to the domain wall motion caused by a spin current.  

\section{Conclusions}

In conclusion, we proposed a reprogrammable magnonic hardware architecture on a single magnetic wafer based on two types of waveguides using domain walls and EASA stripes. Utilizing the chiral property due to DMI in the domain wall wire, we demonstrated by micromagnetic simulations the first building block on this spin wave architecture --- a spin wave diode. Our findings open the gate towards reprogrammable pure spin wave circuits on a single magnetic wafer and ultimately realizing a magnonic computer.

\section{Methods}

The simulation is performed in COMSOL Multiphysics using the mathematical module where the LLG equation is transformed into weak form and solved by the generalized-alpha method (amplification of high frequency is $0.6$) in a 3-dimensional environment. The sample is a yttrium iron garnet (YIG) thin film (size 1600nm $\times$ 1600nm $\times$ 30nm if not mentioned otherwise). The parameters for YIG are: the anisotropy of the magnetic wafer $K = 3.88 \times 10^4$ A/m, exchange coefficient $A_{\rm ex} = 3.28 \times 10^{-11}$ A$\cdot$m, the gyromagnetic ratio $\gamma = 2.21 \times 10^5$ Hz/(A/m), \cite{yan_all-magnonic_2011} the DMI coefficient $D = 1.0 \times 10^{-3}$ A. The thickness of the EASA layer is 3nm and $K_s = 1.0\times 10^{-3}$ A. In order to stabilize the domain wall, a hard-axis anisotropy in $z$ direcion $K_h = 1.0 \times 10^5$ A/m is applied and a groove with thickness of 10nm is made to pin the domain wall. The frequency of the exciting field $f = 2$ GHz is applied locally at the position indicated by the green bar in each simulation figures. The damping coefficient in the working area $\alpha = 10^{-4}$, while the damping near the boundary is set to $\alpha = 0.5$ to eliminate boundary effects such as reflections. In the simulation of \Figure{fig:swdp}, the additional perpendicular anisotropy in the terminal area is taken as $K_s = 0.8K$. 

\bigskip

{\bf Acknowledgements:} J.X. thanks Yizheng Wu, Donglai Feng, and Lei Zhou for helpful remarks on the manuscript. This work was supported by the National Natural Science Foundation of China (11474065, 91121002), National Basic Research Program of China (2014CB921600, 2011CB925601, and 2015CB921400). R.W. also acknowledges support of the 1000-talent program. Work at UCI was supported as part of the SHINES, an Energy Frontier Research Center funded by the U.S. Department of Energy, Office of Science, Basic Energy Sciences under Award No. SC0012670. 

%{\bf Author Contributions:} J.L. \& J.X. did the analytical calculations. W.Y. did the micromagnetic simulations. J.X. planned and supervised the study. R.W. supervised J.L. J.X., J.L. and W.Y. wrote the manuscript. All authors discussed the results and worked on the manuscript.  
%
%{\bf Additional Information:} Correspondence and requests for materials should be addressed to J.X.  
%
%{\bf Competing financial interests:} J.L., W.Y. and J.X. have two Chinese patent applications related to this work, number 2015101233200, ``a waveguide structure for unidirectional transport of spin wave'', and number 2015101037421, ``a design of spin wave diode''.  

\bibliographystyle{apsrev} 
%\bibliography{bb}

\end{document}